\documentclass[twocolumn]{aastex61}

\usepackage{amsmath}
\usepackage{multirow}
\usepackage{xspace}
\usepackage{color,soul}

\newcommand{\sneia}{SNe~Ia\xspace}
\newcommand{\snia}{SN~Ia\xspace}
\newcommand{\cbv}{SN~2017cbv\xspace}
\newcommand{\host}{NGC~5643\xspace}
\newcommand{\los}{LOS\xspace} 
\newcommand{\naid}{Na~I~D\xspace}  
\newcommand{\cahk}{Ca~II~H\&K\xspace}
\newcommand{\cak}{Ca~II~K\xspace}
\newcommand{\cah}{Ca~II~H\xspace}
\newcommand{\ki}{K~I\xspace}  
\newcommand{\nai}{Na~I\xspace}  
\newcommand{\caii}{Ca~II\xspace}
\newcommand{\hi}{H~I\xspace}  

\received{}
\revised{}
\accepted{}
\submitjournal{ApJL}

\shorttitle{\cbv high-resolution spectroscopy}
\shortauthors{Ferretti et al.}

\begin{document}

\title{No evidence of circumstellar gas surrounding Type Ia Supernova \cbv}

\correspondingauthor{Raphael Ferretti}
\email{raphael.ferretti@fysik.su.se}

\author{Raphael Ferretti}
\affil{Oskar Klein Centre, Department of Physics, Stockholm University, Albanova, SE 106 91 Stockholm, Sweden\\}

\author{Rahman Amanullah}
\affiliation{Oskar Klein Centre, Department of Physics, Stockholm University, Albanova, SE 106 91 Stockholm, Sweden\\}

\author{Mattia Bulla}
\affiliation{Oskar Klein Centre, Department of Physics, Stockholm University, Albanova, SE 106 91 Stockholm, Sweden\\}

\author{Ariel Goobar}
\affiliation{Oskar Klein Centre, Department of Physics, Stockholm University, Albanova, SE 106 91 Stockholm, Sweden\\}

\author{Joel Johansson}
\affiliation{Benoziyo Center for Astrophysics, Weizmann Institute of Science, 76100 Rehovot, Israel\\}
\affiliation{Department of Physics and Astronomy, Division of Astronomy and Space Physics, Uppsala University, Box 516, SE 751 20 Uppsala, Sweden\\}

\author{Peter Lundqvist}
\affiliation{Oskar Klein Centre, Department of Astronomy, Stockholm University, Albanova, SE 106 91 Stockholm, Sweden\\}



\begin{abstract}

Nearby type Ia supernovae (\sneia), such as \cbv, are useful events to address the question of what the 
elusive progenitor systems of the explosions are. \citet{2017ApJ...845L..11H} suggested that the early 
blue excess of the lightcurve of \cbv could be due to the supernova ejecta interacting with a 
nondegenerate companion star. 
Some \snia progenitor models suggest the existence of circumstellar (CS) environments 
in which strong outflows create low density cavities of different radii.
Matter deposited at the edges of the cavities, should be at distances at which photoionisation 
due to early ultraviolet (UV) radiation of \sneia
causes detectable changes to the observable \naid and \cahk absorption lines.
To study possible narrow absorption lines from such material, 
we obtained a time-series of high-resolution spectra of \cbv 
at phases between $-14.8$ and $+83$ days with respect to $B$-band maximum, 
covering the time at which photoionisation is predicted to occur. 
Both narrow \naid and \cahk are detected in all spectra, 
with no measurable changes between the epochs. 
We use photoionisation models to rule out the presence of \nai and \caii gas clouds 
along the line-of-sight of \cbv between 
$\sim8\times10^{16}$--$2\times10^{19}$~cm and \caii within $\sim10^{15}$--$10^{17}$~cm, respectively.
Assuming typical abundances, the mass of a homogenous spherical CS gas shell with radius $R$ must be limited to 
$M^{\rm CSM}_{\rm \hi}<3\times10^{-4}\times(R/10^{17}[{\rm cm}])^2$~M$_{\sun}$.
The bounds point to progenitor models that deposit little gas in their CS environment.
\end{abstract}

\keywords{supernovae: individual (\cbv)} 



\section{Introduction} \label{sec:intro}

Type Ia supernovae (\sneia) 
are of great importance to modern cosmology, 
because they are standardisable candles 
\citep[see e.g.][for a review]{2011ARNPS..61..251G}.
Although thousands of \sneia have been observed, 
the physics of the progenitor system leading to the explosions is not fully understood.
There are two prevalent progenitor models for \sneia, both of which have 
some observational support.
The models involve the thermonuclear explosion of a carbon-oxygen (C/O) white dwarf (WD)
in a binary system with another star, which it merges with or accretes mass from.
If the companion star is another WD, 
the system is referred to as a double degenerate \citep[DD,][]{1984ApJS...54..335I,1984ApJ...277..355W},
and if it is a main sequence or giant star, a single degenerate progenitor \citep[SD,][]{1973ApJ...186.1007W}.
More complicated systems, such as common envelope (or symbiotic) binaries \citep{2012Sci...337..942D},
and colliding WDs have also been proposed \citep{2015MNRAS.454L..61D}.

The circumstellar (CS) environment of \sneia should hold clues to the progenitor systems.
SD progenitors for instance are believed to have strong outflows, which excavate low density cavities into 
the surrounding interstellar medium (ISM) and deposit matter at the edges \citep{2007ApJ...662..472B}.
Similarly, DD Helium+C/O binary systems should create cavities with smaller radii \citep{2013ApJ...770L..35S}.
On much smaller scales, tidal effects in DD progenitors can deposit matter into the CS medium \citep{2013ApJ...772....1R}.

Strong upper limits on outflowing matter have been set with radio \citep{2016ApJ...821..119C,2017ApJ...842...17K} and
X-ray \citep{2014ApJ...790...52M} observations of individual \sneia.
Furthermore, the lack of thermal emission in mid- and far-infrared wavelengths, has set strong limits on 
the presence of CS dust \citep{2013MNRAS.431L..43J,2017MNRAS.466.3442J}.
Nevertheless, observations such as predominately blueshifted profiles 
of narrow \naid absorption lines point to outflowing material  
along the lines-of-sight (\los) \citep{2011Sci...333..856S,2013MNRAS.436..222M}.
However, the blueshifted profiles and frequently observed large \nai column densities \citep{2013ApJ...779...38P}
could also be explained by 
desorption from ISM dust grains, when they are exposed to the radiation of \sneia \citep{2014MNRAS.444L..73S}.

Along with a recent method, which follows variable reddening of \sneia \citep{2018MNRAS.473.1918B}, 
variations in narrow absorption line profiles can be used to locate gas close to \sneia. 
Before maximum brightness, photoionisation should lead to
a decrease of characteristic absorption lines \citep{2009ApJ...699L..64B}.
At later phases recombination of the same gas could lead to the increase in the same absorption lines. 
Variations due to photoionisation can be used to determine the distance of the gas from a supernova.
However, geometric effects \citep{2010A&A...514A..78P} and changing levels 
of foreground light \citep{2016ApJ...816...57M} can also lead to similar variations.

A small number of \sneia with variable absorption lines have been observed to date:
\begin{itemize}
\item SN~2006X \citep{2007Sci...317..924P} showed a changing \naid profile at late times, which could point 
to recombination or geometric effects \citep{2008AstL...34..389C}. 
\item SN~2007le \citep{2009ApJ...702.1157S} showed an increasing \naid component.
\item SN~2011fe \citep{2013A&amp;A...549A..62P} showed slight variations, consistent with geometric effects.
\item PTF11kx \citep{2012Sci...337..942D}, a peculiar supernova which showed many variable absorption features and is
believed to have had a symbiotic progenitor.
\item SN~2013gh \citep{2016A&amp;A...592A..40F} had a varying \naid component consistent 
with photoionisation.
\item SN~2014J \citep{2015ApJ...801..136G} showed a varying \ki line while \naid remained unchanged, 
which is consistent with photoionisation, although \citet{2016ApJ...816...57M} argue that the gas 
is unlikely to have been CS matter.
\end{itemize}
A larger sample of 14 \sneia with multi-epoch high-resolution spectra \citep{2014MNRAS.443.1849S}, 
did not reveal further examples of varying absorption lines.
However, almost all existing time-series are taken at too late phases 
to probe for photoionisation of CS gas \citep[see ][]{2016A&amp;A...592A..40F}.

Because of its low redshift, early discovery and a \los seemingly clear of an ISM, 
the recently discovered \cbv \citep{2017ApJ...845L..11H}
presents a good opportunity to search for photoionisation of CS gases.
Below we present the high-resolution spectra we obtained (Section~\ref{sec:obs}), 
in which we identify narrow \naid and \cahk absorption features and search for variations in them (Section~\ref{sec:feat}).
Using a simple photoionisation model, we then set limits on the presence of CS gases (Section~\ref{sec:phot}).
Finally, we compare our observations to the findings of \citet{2017ApJ...845L..11H} 
and conclude (Section~\ref{sec:disc}~\& \ref{sec:conc}).

\section{Observations}
\label{sec:obs}

\begin{table*}
       	\centering
  	\begin{tabular}{c l@{\,}r c c c c c c}
    	\hline\hline
    	{ MJD} & \multicolumn{2}{c}{ UT Date}  &
	{ Exp. time} & { Set-up} &   { Epoch} &  { $R$ $(\lambda/\delta \lambda)$} & { S/N$^{\dagger}$} & { H$_2$O column}\\
	& \multicolumn{2}{c}{ } & { (s)} & & { (days)} & & & { (mm)}\\
    	\hline
	{ 57826.3} & { Mar.}&{ 14.3} & { $1800$} & { 1.0" DIC1 390+580} & { -14.8} & { 62,000} & { 110} & { $13.6\pm0.3$}\\
	{ 57831.2} & { Mar.}&{ 19.2} & { $2 \times 1700$}  & { 0.8" DIC1 390+580} & { -9.9} & { 66,000} & { 150} & { $6.3\pm0.1$}\\
	{ 57846.1} & { Apr.}&{ 03.1} & { $2 \times 2000$} & { 0.8" DIC1 390+580} & { 5.0} & { 52,000} & { 210} & { $1.4\pm0.1$}\\
	{ 57923.2} & { Jun.}&{ 19.2} & { $2 \times 1700$} & { 0.8" DIC1 390+580} & { 82.1} & { 56,000} & \multirow{2}{*}{ $\bigg \}$110} & { $1.2\pm0.2$}\\
	{ 57924.0} & { Jun.}&{ 20.0} & { $2 \times 1700$} & { 0.8" DIC1 390+580} & { 82.9} & { 63,000} & & { $2.1\pm0.2$}\\
	\hline\hline
	\multicolumn{5}{l}{ {$^{\dagger}$around $5900$ \AA}}\\
  	\end{tabular}
  	\caption{The obtained UVES spectra. Epochs are with respect to the $B$-band maximum on MJD 57841.1 \citep{2017ApJ...845L..11H}. 
	Resolutions are inferred from the full-width-half-maxima of several telluric lines and the H$_2$O columns were computed with Molecfit.
	\label{tab:spec}}
\end{table*}

\cbv was discovered on March 10 UT (MJD 57822.14) by the Distance Less Than 40 Mpc (DLT40) 
supernova survey \citep{2017ATel10158....1T} and subsequently classified as a young \snia \citep{2017ATel10164....1H}.
The supernova is located in the outskirts of \host at 
z $=0.003999(7)$ \citep{2004AJ....128...16K}, 
at $\alpha = 14^{h}32^{m}34.42^{s}, \delta = -44^{\circ}08'02.8"$ (J2000),
a \los with galactic extinction $E(B-V)_{MW}=0.15$~mag \citep{2011ApJ...737..103S}.
Lightcurve analysis by \citet{2017ApJ...845L..11H} determined that \cbv peaked at
MJD 57841.07 in $B$-band with $\Delta m_{15}(B)=1.06$ mag.

Because \cbv was a good candidate to search for CS gas,
we triggered our ESO TOOs 098.A-0783(A) and 098.A-0783(B) (P.I. Amanullah)
to obtain spectra with the 
Ultraviolet and Visual Echelle Spectrograph  \citep[UVES;][]{2000SPIE.4008..534D}  
on UT2 at the Very Large Telescope (VLT).
We reduced the spectra using the REFLEX (ESOREX) reduction pipeline
provided by ESO \citep{2010SPIE.7737E..28M} and used the telluric line correction
software Molecfit \citep{2015A&amp;A...576A..77S,2015A&amp;A...576A..78K} where necessary.
The obtained spectra are summarised in Table~\ref{tab:spec}.

The first UVES spectrum was taken on MJD 57826.3, 4.2 days after discovery,
at an epoch of $-14.8$~days before $B$-band maximum.
Two follow-up spectra were obtained bracketing maximum light 
to cover the time frame at which changes due to photoionisation could be expected.
Finally, two late spectra were taken on back-to-back nights to cover phases 
during which late-time absorption line variations have been observed in the past \citep[e.g. SN~2006X,][]{2007Sci...317..924P}.
In the following, the two last spectra are treated as one epoch, since \sneia 
only evolve slowly at late phases and an improved signal-to-noise ratio (S$/$N) is 
achieved by coadding them.

\section{Narrow absorption features}
\label{sec:feat}

We searched all spectra for narrow absorption features in the rest frame of \host. 
While \cahk lines were immediately apparent in all spectra, \naid absorption was only revealed after applying
telluric line corrections. 
We searched for, but did not detect CH and CH$+$ or any diffuse interstellar bands (DIBs) 
such as those identified in \citet{2005A&amp;A...429..559S}.
The frequently studied DIBs at $\lambda\lambda$ 5780 and 5797 fall in between the spectral 
arms of the UVES spectra and \ki lies outside the range of the instrument with 
the chosen set up. 

\begin{figure*}
   \resizebox{\hsize}{!}
            {\includegraphics{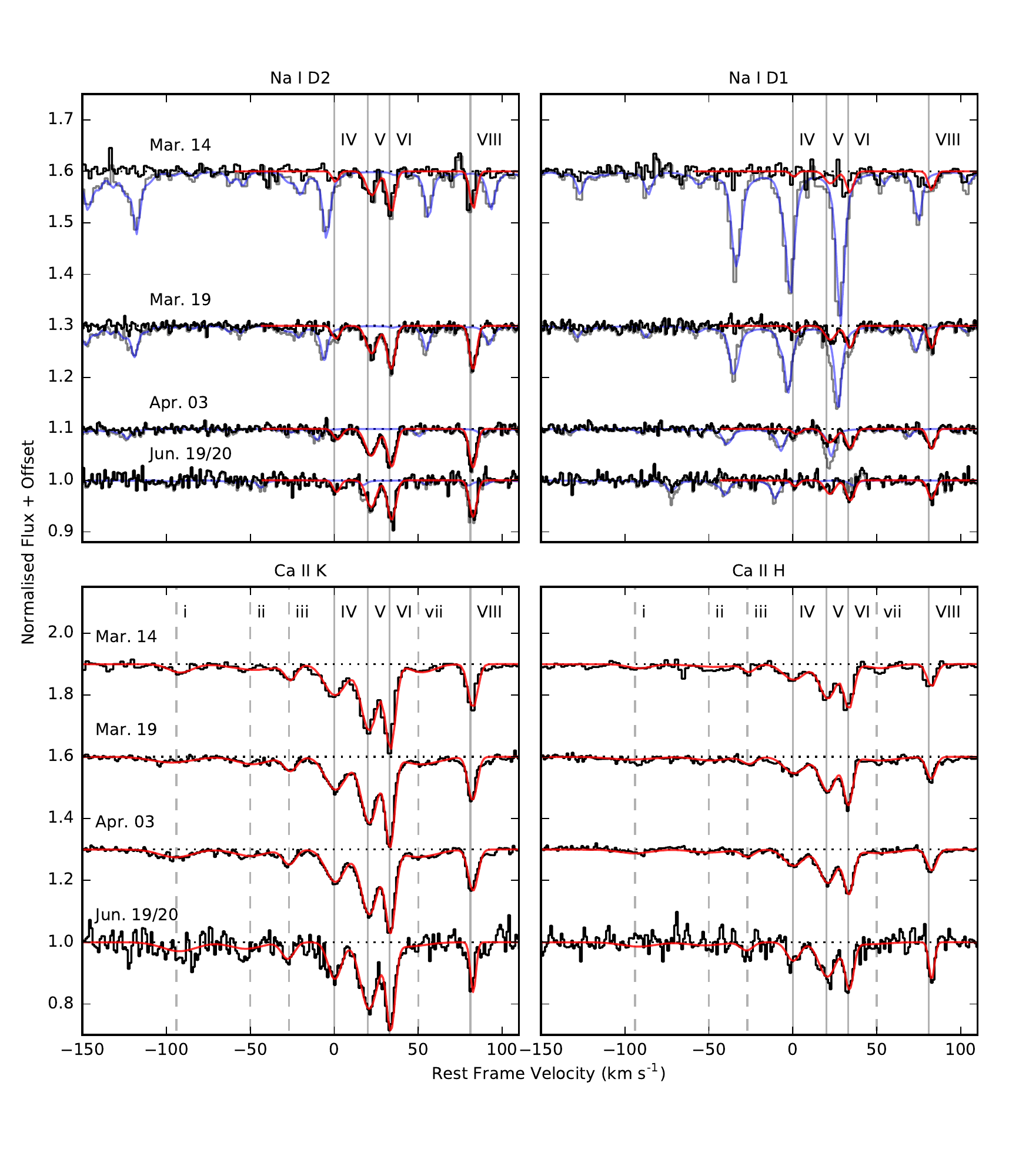}}
      \caption{The detected \naid and \cahk features in velocity space of the host galaxy rest frame at z $=0.004$.
      For clarity, the spectra, which are shown in black, have been normalised, offset and, in the case of \naid, telluric line corrected.
      The uncorrected spectra are shown in grey along with the telluric line model in light blue.
      The fitted Voigt profiles are plotted in red and vertical lines indicate individual line components.
      Features present in both \naid and \cahk are marked with solid grey lines and labeled with capital 
      roman numerals, while features only detected in \cahk with dashed grey lines and lower case roman numerals.
              }
         \label{fig:feat}
   \end{figure*}

Around z~$=0.004$, many telluric lines can make the identification 
of \naid difficult. In fact the telluric features in the first two epochs 
are more prominent than \naid.
In Table~\ref{tab:spec} the H$_2$O column determined from telluric line fitting is shown for reference.
The \naid features could be identified because,
\begin{itemize}
\item several features remained after telluric line correction at the redshift of \host,
\item they appeared as doublets with the separation and line ratios characteristic of \naid,
\item they did not shift with the telluric lines between epochs,
\item and they appeared to have the same rest frame velocity as the deepest features of \cahk.
\end{itemize}
The detected absorption lines are plotted in Figure~\ref{fig:feat} after normalising the continua 
and correcting for telluric features. 

We visually compared the profiles of each epoch for differences.
In the first epoch a small peak appears in next to the most redshifted feature 
in \naid\/2 (labeled VIII in Figure~\ref{fig:feat}).
Notably, there is no feature at the corresponding wavelength in \naid\/1 
and several comparable outliers can be identified in the continuum of this spectrum. 
For this reason we ignore the peak in the further analysis. 
No other significant differences could be visually identified between the epochs.

To search for absorption line variations, 
we measured the pixel-for-pixel equivalent widths ($W$) of each epoch.
Where possible, we measured $W$ of individual features or,
if the lines are blended, across groups of features.
In neither case significant trends could be identified.
The average total values measured are $W_{\rm \naid\/1}=13.9\pm1.4$,
$W_{\rm \naid\/2}=31.2\pm1.3$,
$W_{\rm \cah}=69.3\pm2.5$ and 
$W_{\rm \cak}=123.5\pm2.5$~m\AA.

We further fit Voigt profiles to the features using VPFIT\footnote{\url{http://www.ast.cam.ac.uk/~rfc/vpfit.html}}, 
whereby we fit the profiles of 
each doublet of the same absorber simultaneously (\naid\/1 with D2 and \cah with K). 
To establish the best over all profile, we initially fit all epochs simultaneously.
The \cahk profiles can be fit well by eight individual components, while 
four of those match the profile of \naid.
In Table~\ref{tab:cd}, the Voigt profile parameters, redshift $(z)$, column density $(N_X)$,  and Doppler widths $(b_X)$,
of the simultaneous fits are presented, 
whereby the features are labeled the same way as in Figure~\ref{fig:feat}.

The most prominent features IV--VI and VIII have close enough redshifts in
both \naid and \cahk to correspond to the same gas clouds along the \los to \cbv.
Interestingly, $b_{\rm \caii}$ of these features appear to be 
significantly larger than $b_{\rm \nai}$.
This suggests either that there are more unresolved features in \cahk,
or \caii is depleted in the colder regions of these gas clouds.
The ratios $({\rm N_{\caii}/N_{\nai}})$ of these features 
are also shown in Table~\ref{tab:cd}.

\begin{figure*}
   \resizebox{\hsize}{!}
            {\includegraphics{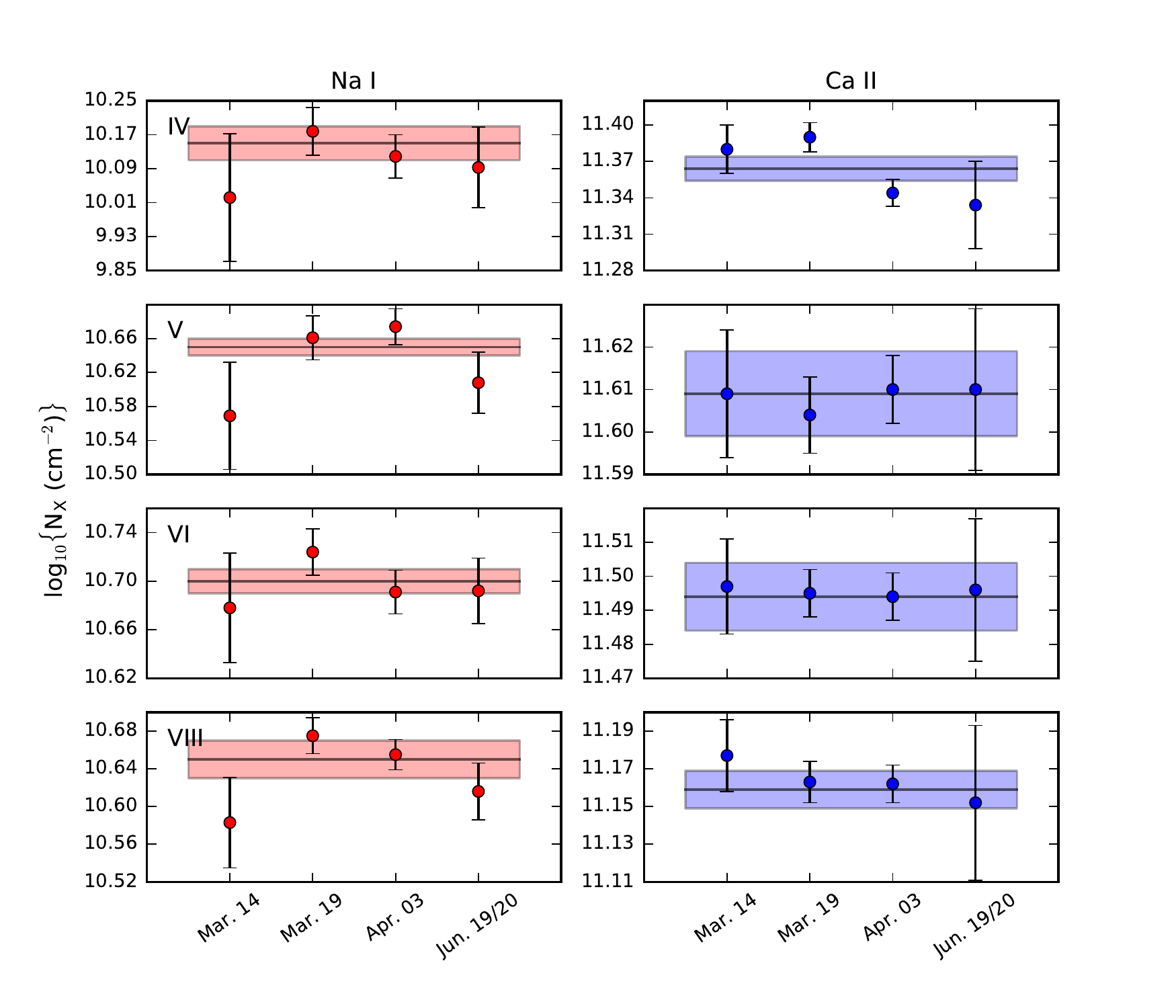}}
      \caption{
      Column densities of the four most prominent absorption features IV, V, VI and VIII inferred from the Voigt profile fits.
      The data points are computed from each epoch individually with $1\sigma$ error-bars, 
      while the horizontal lines are determined by the simultaneous fit of all epochs with a $1\sigma$ band.
              }
         \label{fig:cd}
   \end{figure*}

To search for time dependence, we fit the Voigt profiles to each epoch individually.
The profiles are plotted in red over the spectra in Figure~\ref{fig:feat}.
We studied $N_{\rm X}$ and $b_{\rm X}$ of each feature 
without finding any significant trends. 
In Figure~\ref{fig:cd} it can be seen that $N_{\rm X}$ of the most prominent features are consistent with the 
values determined from the simultaneous fits of all epochs.

 \begin{table*}
  \centering
  \begin{tabular}{c c c r@{\,}l r@{\,}l c r@{\,}l r@{\,}l c}
    \hline\hline
    { Feature} & { $v^{\dagger}$} & {$z_{\rm \nai}$} & \multicolumn{2}{c}{ $\log_{10}\{N_{\rm \nai}\}$} & \multicolumn{2}{c}{$b_{\rm \nai}$} 
    & {$z_{\rm \caii}$} & \multicolumn{2}{c}{ $\log_{10}\{N_{\rm \caii}\}$} & \multicolumn{2}{c}{$b_{\rm \caii}$} & {$N_{\rm \caii}/N_{\rm \nai}$} \\
    &  { (km s$^{-1}$)} & & \multicolumn{2}{c}{ (cm$^{-2}$)} & \multicolumn{2}{c}{ (km s$^{-1}$)} & & \multicolumn{2}{c}{ (cm$^{-2}$)} 
    & \multicolumn{2}{c}{ (km s$^{-1}$)} & \\ 
    \hline
    	{ i } & { -94} & {--} & \multicolumn{2}{c}{ --} & \multicolumn{2}{c}{ --} & {0.003684(2)} & { 10.92} & { $\pm$ 0.03} & { 14.1} & { $\pm$ 1.0} & { -- } \\
	{ ii }  & { -50} & {--} & \multicolumn{2}{c}{ --} & \multicolumn{2}{c}{ --} & {0.003834(3)} & { 10.85} & { $\pm$ 0.03} & { 12.3} & { $\pm$ 1.2} & { -- } \\
	{ iii }  & { -27} & {--} & \multicolumn{2}{c}{ --} & \multicolumn{2}{c}{ --} & {0.003911(1)} & { 10.83} & { $\pm$ 0.02} & { 4.6} & { $\pm$ 0.5} & { -- } \\
	{ IV }  & { 1} & {0.004005(1)} & { 10.15} & { $\pm$ 0.03} & { 1.7} & { $\pm$ 0.7} & {0.004002(1)} & { 11.37} & { $\pm$ 0.01} & { 7.8} & { $\pm$ 0.2} & { 17.4}\\
	{ V }  & { 21} & {0.004074(1)} & { 10.66} & { $\pm$ 0.01} & { 3.5} & { $\pm$ 0.3} & {0.004070(1)} & { 11.60} & { $\pm$ 0.01} & { 5.9} & { $\pm$ 0.1} & { 8.9}\\
	{ VI }  & { 33} & {0.004114(1)} & { 10.72} & { $\pm$ 0.01} & { 1.5} & { $\pm$ 0.3} & {0.004112(1)} & { 11.50} & { $\pm$ 0.01} & { 2.4} & { $\pm$ 0.1} & { 6.2}\\
	{ vii }  & { 50} & {--}  & \multicolumn{2}{c}{ --} & \multicolumn{2}{c}{ --} & {0.004168(4)} & { 10.98} & { $\pm$ 0.04} & { 15.1} & { $\pm$ 1.5} & { -- } \\
	{ VIII }  & { 82} & {0.004277(1)} & { 10.67} & { $\pm$ 0.02} & { 0.7} & { $\pm$ 0.4} & {0.004276(1)} & { 11.16} & { $\pm$ 0.01} & { 2.3} & { $\pm$ 0.2} & { 3.2}\\

    \hline\hline
\multicolumn{10}{l}{ $^{\dagger}$ inferred from $z_{\rm \caii}$ with respect to $z = 0.003999$} 
  \end{tabular}
  \caption{Redshifts $z_X$, Column densities ${\rm N_X}$ and Doppler widths of ${\rm b_X}$ of the fitted Voigt profile components. 
  The values are the average parameters obtained from fitting the profiles of each spectrum.
  Features are labeled the same way as in Figure~\ref{fig:feat}, where
  capital roman numerals correspond to features detected in both \naid and \cahk, 
  while the lower case roman numerals correspond to features only detected in \cahk.
  \label{tab:cd}}
\end{table*}

\section{Photoionisation}
\label{sec:phot}

       \begin{figure*}
   \resizebox{\hsize}{!}
            {\includegraphics{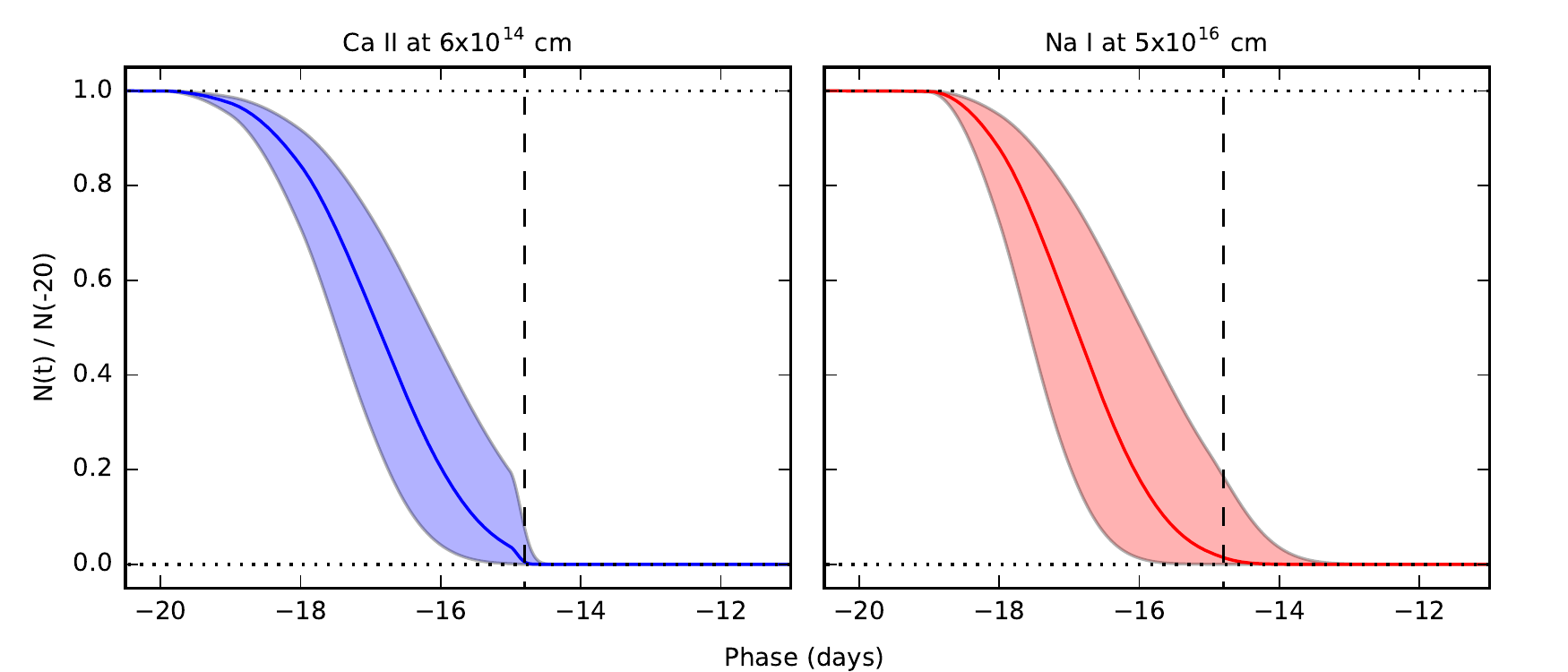}}
      \caption{ \nai and \caii ionisation curves at the inner exclusion limit. Any gas clouds closer to a \snia than those shown are
      ionised before an epoch of $-14.8$ days. 
      The bands show the changes in the ionisation faction if the UV flux of the SN varies by $\pm1$~mag (factor $2.5$).
              }
         \label{fig:N}
   \end{figure*}

We use a photoionisation model described in \citet{2009ApJ...699L..64B}, 
in which we assume a homogenous thin shell of CS gas that is optically thin to ionising photons.
The fraction of such a gas $X$ in a shell of radius $R$ that is ionised at a given time is
\begin{equation}
I_X(t;R)=\frac{D^2}{R^2}\int_{0}^{\lambda_{I}}\alpha_X(\lambda)\frac{f_{\lambda}(\lambda,t)}{hc}\lambda\,{\rm d}\lambda,
\end{equation}
where $f_{\lambda}(\lambda,t)$ is the spectral energy distribution (SED) of a supernova at a distance $D$ from the observer, $\alpha_X(\lambda)$ 
is the wavelength dependent ionisation cross-section\footnote{Obtained via \url{phidrates.space.swri.edu}} \citep{2015P&amp;SS..106...11H}
and $\lambda_I$ the ionisation energy of the gas. 
The column density $N_X(t)$ is then defined by 
\begin{equation}
\frac{{\rm d}N_X(t)}{{\rm d}t}=-N_X(t)I_X(t;R),
\end{equation}
which implies that
\begin{equation}
N_X(t';R,N_0)=N_0{\rm exp}\left\{-\int_{t_0}^{t'}I_X(t;R){\rm d}t\right\}.
\end{equation}
We use the SED from \citet{2015MNRAS.453.3300A} based on SN~2011fe, since this supernova has the best temporal coverage of high S$/$N ultraviolet (UV) spectra.
$N_X(t)$ then depends on two free parameters, the initial column density $N_0$ and $R$.

In the absence of absorption line variations, the model can be used to set limits
on $R$ and $N_0$. 
For a given set of observations, one can define a range $R_X^{\rm excl.}$, within which any gas would be ionised.
Thereby $R_X^{\rm inner}$ defines the radius at which all gas is ionised before the first spectrum is obtained and 
$R_X^{\rm outer}$ is the radius beyond which the ionising flux is negligible.
To obtain a strong inner radius limit, the first spectrum must be obtained as early as possible. 

The spectral coverage of \cbv implies that $R^{\rm excl.}_{\rm \nai}\approx5\times10^{16}$--$2\times10^{19}$~cm
and $R^{\rm excl.}_{\rm \caii}\approx6\times10^{14}$--$10^{17}$~cm.
In Figure~\ref{fig:N} the fractional ionisation curves at $R_{\rm \nai}^{\rm inner}$ and $R_{\rm \caii}^{\rm inner}$ are shown.
Since the ultraviolet (UV) flux of \sneia can vary considerably between supernovae, we consider a $\pm1$~mag (factor 2.5) range in UV flux. 
If the SED underestimates the UV flux of \cbv by this factor, $R^{\rm inner}_{\rm \nai}\approx8\times10^{16}$~cm
and $R^{\rm inner}_{\rm \caii}\approx10^{15}$~cm.

One can furthermore define $N_{X}^{\rm upper}$, the highest column density a gas cloud within 
the excluded radius range can have to not be detected.
In the cases of \nai and \caii gas, one can require that an absorption feature must be identifiable in both profiles of the 
\naid and \cahk.
This implies it must be detectable in \naid\/1 and \cah, 
which have the weaker oscillator strengths of their respective doublets.

Assuming a feature with a Gaussian profile and a full-width-half-maximum of $0.1$~\AA, the column density would
need to be more than ${\rm log}_{10}\{N_{\rm \nai}^{upper}\,[cm^{-2}]\}=10.4$ for a line to be identifiable above 
the noise in \naid\/1. 
Notably, feature IV has a lower column density than this limit and is not detected in \naid\/1 above the noise of the first epoch.
Only in the later spectra, with higher S$/$N and lower water column, 
feature IV can be identified.
Using the same criterion, 
a column density greater than ${\rm log}_{10}\{N_{\rm \caii}^{upper}\,[cm^{-2}]\}=10.7$
is necessary to identify a feature in \cah in the first epoch.

One can further determine a upper bound on the enclosed mass of CS gas from the upper column density limits.
Assuming a homogenous shell of gas with a radius $R$, 
$M^{\rm CSM}_{\rm \nai}<6\times10^{-11}\times(R/10^{17}[{\rm cm}])^2$~M$_{\sun}$ and 
$M^{\rm CSM}_{\rm \caii}<2\times10^{-10}\times(R/10^{17}[{\rm cm}])^2$~M$_{\sun}$, where the values are scaled to $R=10^{17}$~cm.
For typical abundances of $N_{\rm \nai}/N_{\rm \hi}\sim10^{-8}$ \citep{2000ApJ...544L.107W},
this suggests an \hi mass of $M^{\rm CSM}_{\rm \hi}<3\times10^{-4}\times(R/10^{17}[{\rm cm}])^2$~M$_{\sun}$.
If the gas has been swept up from the ISM surrounding the progenitor, the \hi volume density must have been $<10^{2}\times((10^{17}[{\rm cm}])/R)$~cm$^{-3}$.

After photoionisation, gas is expected to recombine, whereby the recombination rate is dependent on the 
temperature and electron density of the gas cloud.
To describe recombination on the time scale of the variations observed in SNe~2006X and 2007le, 
high electron densities must be assumed \citep[e.g. $10^7$~cm$^{-3}$,][]{2009ApJ...702.1157S}.
This has lead to alternative explanations through geometric effects \citep{2008AstL...34..389C} 
or photon-induced desorption \citep{2014MNRAS.444L..73S}.
The spectra of the last epoch of \cbv were taken at similar time-scales as the examples above,
but show no comparable variations. 

\section{Discussion}
\label{sec:disc}

The observations suggest
that the detected absorption lines stem from gas at interstellar distances.
In \cahk, the gas clouds span a velocity range of $>170$~km~s$^{-1}$,
a dispersion that would be unusual for the disc of a spiral galaxy at the projected distance of $11.6$~kpc from  
the core.
Thus some of the gas clouds might be located in the halo of \host.
In the Milky Way, it has been found that the ratio $N_{\rm \caii}/N_{\rm \nai}$ is lower in 
gas clouds in the disc than in the halo \citep{1975ApJ...198..545C}.
This relation has been used to distinguish between disc and halo absorbing systems in 
quasar spectra \citep[e.g. ][]{1985MNRAS.216P..41B}.
In \cbv, feature VIII has the lowest  $N_{\rm \caii}/N_{\rm \nai}$ value, consistent with gas from the disc.
Features IV--VI have higher ratios, suggesting that they could be located in the halo.

Assuming similar ISM properties to our Milky Way, 
we can infer reddening from $W_{\rm \naid}$ \citep{2012MNRAS.426.1465P}.
This suggests a negligible $E(B-V)=0.016\pm0.003$~mag, in agreement with \citet{2017ApJ...845L..11H}.

Some \sneia have unusually steep extinction curves \citep[e.g.][]{2015MNRAS.453.3300A},
which could be explained by the presence of CS dust \citep{2008ApJ...686L.103G}.
To affect the extinction curve, the dust must be situated at distances closer than a few $10^{17}$~cm.
Assuming the dust is traced by \nai and \caii gas, our observations exclude 
CS dust at distances of $\sim10^{15}$--$10^{19}$~cm from \cbv.
\citet{2017ApJ...836...13H} propose that the steep extinction curves are the result of dust grains shattered 
by radiation pressure. In line with the suggestions in \citet{2014MNRAS.444L..73S}, this could account for 
the frequently observed blueshifted and unusually deep \naid 
profiles \citep{2011Sci...333..856S,2013MNRAS.436..222M,2013ApJ...779...38P}.

\citet{2017ApJ...845L..11H} propose that the early blue excess 
of \cbv could be due to the supernova ejecta hitting a non-degenerate companion star \citep{2010ApJ...708.1025K}.
The progenitor system containing a subgiant companion star must have sustained strong stellar winds during accretion.
Models suggest that the outflows should excavate large low-density cavities with radii of 
$\sim10^{19}$--$10^{20}$~cm \citep{2007ApJ...662..472B} into the ISM.
Since our observations exclude photoionisation of \nai up to $\sim 2\times10^{19}$~cm from \cbv, 
a cavity larger than this is compatible with \citet{2007ApJ...662..472B}.
However, the smaller cavities with a few $10^{17}$~cm predicted for DD He+C/O progenitors \citep{2013ApJ...770L..35S}
can be excluded.

\section{Conclusions}
\label{sec:conc}

We have detected multiple \nai and \caii gas clouds along the \los to \cbv,
a \snia on the outskirts of its host galaxy \host.
We have obtained multi-epoch high-resolution spectra with UVES 
starting at an early epoch of $-14.8$ days before maximum brightness.
Due to the extensive time coverage of \cbv, we are sensitive to photoionisation occurring 
in gas clouds over a large part of the CS environment.
We did not find any time-evolution in any of the detected narrow absorption features, 
which implies that no detectable \nai gas clouds could be present
with $\sim8\times10^{16}$--$2\times10^{19}$~cm and \caii within $\sim10^{15}$--$10^{17}$~cm
from the explosion.
The detected gas clouds must therefore be located further from \cbv than the outer limit, while
any gas closer to the explosion than the inner limit would have been ionised before the first spectrum was obtained.

\citet{2017ApJ...845L..11H} suggest that an early blue excess in the lightcurve could be due to 
ejecta hitting a non-degenerate companion star in a SD progenitor system.
However, they do point out that a lack of a corresponding UV bump is in disagreement 
with the models of \citet{2010ApJ...708.1025K} and propose several explanations for the discrepancy.
A SD progenitor is predicted to have excavated large cavities with radii 
of $\sim10^{19}$--$10^{20}$~cm into the surrounding ISM and deposit matter at the edges \citep{2007ApJ...662..472B}.
Our observations thus exclude the presence of significant amounts of matter from parts of this range.
At the same time, no significant amounts of gas could have been at  radii of a few $10^{17}$~cm,
distances at which DD He+C/O progenitor models predict matter to be deposited \citep{2013ApJ...770L..35S}. 
There are thus several possible interpretations for our observations, or combinations of them: 
\begin{itemize}
\item There is no CS gas and the detected absorption features are part of the ISM of \host and unrelated to \cbv.
\item There is a CS cavity within the exclusion range, 
but the \nai and \caii columns present at the edges are below the detection threshold
with ${\rm log}_{10}\{N_{\rm \nai}^{upper}\,[cm^{-2}]\}=10.4$ and \\
${\rm log}_{10}\{N_{\rm \caii}^{upper}\,[cm^{-2}]\}=10.7$. 
Assuming normal abundances, the swept up \hi mass must be
$M^{\rm CSM}_{\rm \hi}<3\times10^{-4}\times(R/10^{17}[{\rm cm}])^2$~M$_{\sun}$, 
suggesting that the ISM had a very low density.
\item The outflowing matter of an SD progenitor system created a cavity larger 
than $2\times10^{19}$~cm \citep[consistent with models of][]{2007ApJ...662..472B} 
and at least some of the detected absorption features correspond to gas deposited at the edges.
\end{itemize}

Our observations should add useful information to the open progenitor question of \cbv and \sneia in general.
A SD progenitor system as that proposed by \citet{2017ApJ...845L..11H} should have a CS environment comparable to 
that described in \citet{2007ApJ...662..472B}.
While we cannot exclude the full range of possible CS gas shells in this model, we also do not find any evidence for 
significant amounts of CS gas. 
This suggests a progenitor model with little to no CS matter for \cbv.
Late time observations, when \cbv reaches a nebular phase should provide further evidence, if there 
was a non-degenerate companion in the progenitor system. 
In the cases of SNe 2011fe and 2014J, nebular spectra provided strong evidence for the absence of a non-degenerate 
companion star \citep{2015A&amp;A...577A..39L}.

\acknowledgments
The authors acknowledge support from the Swedish Research Council 
(Vetenskapsr\aa det) and the Swedish National Space Board.

   \bibliographystyle{aasjournal} 
   \bibliography{cbv} 

\end{document}